\documentclass[aip,apl,preprint,amsmath,superscriptaddress]{revtex4-1}
\input{epsf.sty}
\usepackage{epsfig}
\usepackage{graphicx}
\usepackage{epstopdf} 
\usepackage{amsmath,amsthm,amssymb,latexsym,amsfonts,times,mathrsfs,verbatim,lipsum}
\usepackage{bm}
\usepackage{color}

\begin{document}
\title{Micromagnetic theory of spin relaxation and ferromagnetic resonance in multilayered metallic films}
 \author{Sergey Bastrukov}
 \email{Sergey\_B@dsi.a-star.edu.sg}
 \author{Jun Yong Khoo} 
 \author{Boris Lukiyanchuk}
 \affiliation{Data Storage Institute (DSI), Agency for Science, Technology and Research (A*STAR), 5 Engineering Drive 1, 117608 Singapore}
 \author{Irina Molodtsova}
 \affiliation{Laboratory of Informational Technologies, Joint Institute for Nuclear Research, 141980 Dubna, Russia}

\date{\today}

\begin{abstract}
Spin relaxation in the ultrathin metallic films of stacked microelectronic devices is investigated 
on the basis of a modified Landau-Lifshitz equation of micromagnetic dynamics in which the damping torque is treated 
as originating from the coupling between precessing magnetization-vector and the introduced stress-tensors of intrinsic and extrinsic magnetic anisotropy.
Particular attention is given to the time of exponential relaxation and ferromagnetic resonance linewidth which are derived in analytic form 
from the equation of magnetization energy loss and Gabor uncertainty relation between the full-width-at-half-maximum in resonance-shaped line and 
lifetime of resonance excitation. The potential of developed theory is briefly discussed in the context of recent measurements.
\end{abstract}

\maketitle

\section{Introduction}
An understanding of relaxation processes in ultrathin films of ferromagnetic metals is crucial to 
the design and construction of microelectronic devices\cite{XLO-11,LS-11}, like magnetic random access memory (MRAM) and spatial light modulators (SLM).
The main source of information about relaxation processes has been and still is the ferromagnetic resonance (FMR) measurements 
aimed at revealing the frequency dependence of the full-width-at-half-maximum 
in FMR spectral line\cite{C-06,B-11}. Traditionally, the results of these experiments are treated within the framework of 
 the phenomenological Landau-Lifshitz-Gilbert  model\cite{K-07,S-09,B-12}, describing the FMR response in terms 
of uniform precession of magnetization ${\bf M}(t)$ with the preserved in time magnitude $|{\bf M}(t)|=M_s$ where $M_s$ is 
the saturation magnetization. The dynamical equation governing precessional motions of ${\bf M}(t)$ can be 
conveniently written in the following general form
 \begin{eqnarray}
\label{e1.1}
{\dot {\bf M}(t)}=-\gamma\mu_0 {\bf T}(t)-{\bf R}(t),
\end{eqnarray} 
where $\gamma$ is the electronic gyromagnetic ratio and $\mu_0$ is the magnetic permeability of free space; SI units are used throughout this paper. 
The vector-function, ${\bf T}(t)=[{\bf M}(t)\times{\bf H}]$, represents the magnetic torque  that drives the free Larmor precession of ${\bf M}(t)$  
about the axis of the dc magnetic field,  ${\bf H}=\mbox{constant}$, in the process of which the Zeeman magnetization energy, $W_m(t)=-\mu_0 {\bf H}\cdot {\bf M}(t)$, is conserved: ${\dot W}_m(t)=0$. Central to understanding the relaxation process is the relaxation function ${\bf R}(t)$ defining the 
rate of magnetization energy loss 
\begin{eqnarray}
\label{e1.2}
&& {\dot W}_m(t)=-\mu_0 {\bf H}\cdot{\dot {\bf M}(t)}=\mu_0{\bf H}\cdot{\bf {\bf R}}(t).
\end{eqnarray}
In this work we focus on Landau-Lifshitz (LL) form of this function ${\bf R}_{in}(t)=\lambda_{in}[{\bf M}(t)\times[{\bf M}(t)\times{\bf H}]]$
which provides geometrically transparent insight into the magnetization-vector motion in the process of aligning 
${\bf M}$ with ${\bf H}$. The material-dependent parameter $\lambda_{in}$ can be  thought of as describing 
the strength effect of the intrinsic anisotropy on the relaxation dynamics of magnetization precession which is constrained by 
the conditions
${\bf M}(t)\cdot {\dot {\bf M}}(t)=0$ and ${\bf M}(t)\cdot {\bf R}(t)=0$. In this paper we consider an alternative 
micromagnetic treatment of ${\bf R}(t)$ according to which the origin of the damping torque responsible for spin relaxation in multilayered metallic films 
is attributed to the coupling between the uniformly precessing magnetization-vector and the stress-tensors of intrinsic and extrinsic magnetic anisotropy.

\section{Stress-tensor representation of micromagnetic damping torque}
 The equilibrium magnetic anisotropy exhibited in the  
 easy and hard axes of magnetization direction~\cite{J-96} is a hallmark of ultrathin films of ferromagnetic metals.  
 Viewing this property from the perspective of the macroscopic electrodynamics of magnetic continuous media~\cite{LLP-84,M-88}, 
 it seems quite natural to  invoke the stress-tensor description of magnetic anisotropy, namely, 
 in terms of symmetric tensors of magnetic-field-dependent stresses. In so doing we adopt the following definition of stress-tensor of intrinsic anisotropy $\sigma^{in}_{lk}$ (generic to both monolithic and multilayered ferromagnetic films) and stress-tensor of extrinsic anisotropy 
 $\sigma^{ex}_{ik}$ (arising from impurities and imperfections of the film crystalline lattice)  
 \begin{eqnarray}
\label{e2.1}
&& \sigma^{in}_{lk}=\frac{\mu}{2}\left[(M_nH_n)\delta_{kl}-(M_lH_k+M_kH_l)\right],\\
\label{e2.2}
&& \sigma^{ex}_{ik}=\frac{\mu}{2}\left[H^2\delta_{kl}-(H_lH_k+H_kH_l)\right],
\end{eqnarray}  
where $\delta_{ik}$ is the Kronecker symbol and $\mu$ stands for the effective magnetic permeability which is derived from the magnetization curve according to the rule\cite{AH-00}: $\mu=\Delta B/\Delta H$. Hereafter ${\bf H}$ refers to the total (applied and internal effective) field. 
It is worth noting that the above stress-tensor description  
of intrinsic and extrinsic magnetic anisotropy is consistent with the definition of the energy density of magnetic field stored in a ferromagnetic film\cite{TG-08}
\begin{eqnarray}
\label{e2.3}
&& u=\frac{1}{2}{\bf B}\cdot{\bf H},\quad {\bf B}=\mu({\bf H}+{\bf M})
\end{eqnarray}
in the sense that the relation between the stress-tensor of combined intrinsic and extrinsic magnetic anisotropy, 
$\sigma_{lk}=\sigma^{in}_{lk}+\sigma^{ex}_{lk}$, and the energy density $u$ is described by 
\begin{eqnarray}
\label{e2.4}
&& u=Tr[\sigma_{lk}]=\sigma_{ll}=\frac{\mu}{2}(MH+H^2),
\end{eqnarray}
where $Tr[\sigma_{lk}]$ stands for the trace of tensor $\sigma_{lk}$. In what follows we focus 
on the effect of above  magnetic stresses on the precessing magnetization vector whose mathematical treatment 
is substantially relied on the symmetric tensor 
\begin{eqnarray}
\label{e2.5}
\gamma_{ik}=[M_s^2\delta_{ik}-M_iM_k],\quad\gamma_{ik}=\gamma_{ki},
\end{eqnarray}
having, in appearance, some features in common with that for isotropic magneto-striction stresses\cite{ABP-68}.  
It can be verified by direct calculation that the stress-tensor representation of the intrinsic relaxation function 
is identical to the LL relaxation function
\begin{eqnarray}
 \label{e2.6}
 &&R_i^{in}=\frac{2\lambda_{in}}{\mu\,M_s^2}\gamma_{ik}\,M_l\,\sigma^{in}_{kl}=\lambda_{in}[M_i(M_kH_k)-H_iM^2_s],\\
 \nonumber
 &&{\bf R}_{in}=\lambda_{in}[{\bf M}(t)\times[{\bf M}(t)\times{\bf H}]].
\end{eqnarray}
In choosing the above form of the tensor $\gamma_{ik}$ we were guided by previous investigations\cite{JMMM-06} of the damping 
terms in ferro-nematic liquid crystals dealing with the tensor constructions of a similar form. 
For the extrinsic relaxation function,  owing its origin to the coupling of $M_l$ with $\sigma^{ex}_{lk}$,
we use the following stress-tensor representation 
\begin{eqnarray}
\label{e2.7}
R_i^{ex}&=&-\frac{\lambda_{\rm ex}}{\mu\Omega}\gamma_{ik}M_l\sigma^{ex}_{lk}\\
\nonumber
&=&-\lambda_{\rm ex}\frac{(M_nH_n)}{\Omega}[M_i(M_kH_k)-H_iM^2_s].
\end{eqnarray} 
The minus sign means that extrinsic damping torque counteracts the damping torque originating from the intrinsic stresses.    
The vector form of extrinsic relaxation function (\ref{e2.7}) reads 
\begin{eqnarray}
\label{e2.8}
{\bf R}_{ex}(t)=-\lambda_{ex}\frac{({\bf M}(t)\cdot {\bf H})}{\Omega}\,[{\bf M}(t)\times[{\bf M}(t)\times {\bf H}]]. 
\end{eqnarray}
As is shown below, the parameter-free frequency of the transient magnetization configuration 
\begin{eqnarray}
\label{e2.9} 
&&  \Omega=\frac{\omega}{1-(\omega_M/\omega)^{1/2}},\,\,\omega_M=\gamma\mu_0M_s,\,\,\omega=2\pi f,
\end{eqnarray}
provides correct physical dimension of the extrinsic damping torque 
and proper account for the empirical dependence of the FMR linewidth $\Delta H$ upon the resonance frequency $f$. 
Making use of argument of physical dimension it is easy to show that  
the material-dependent parameters $\lambda_{in}>0$ and $\lambda_{ex}>0$
(measuring strength of intrinsic and extrinsic stresses on the relaxation process in multilayered film) 
can be represented in terms of dimensionless damping constants  $\alpha$ and $\beta$ (whose magnitudes are  
deduced from the empirical frequency dependence of 
FMR linewidth) as follows  
 \begin{eqnarray}
\label{e2.10} 
&& \lambda_{in}=\alpha\frac{\gamma\mu_0}{M_s},\quad \lambda_{ex}=\beta\left(\frac{\gamma\mu_0}{M_s}\right)^2.
\end{eqnarray}
The net outcome of the above outlined procedure of computing the combined (intrinsic plus extrinsic) damping torque 
 \begin{eqnarray}
\label{e2.10A} 
{\bf R}&=&{\bf R}_{in}+{\bf R}_{ex}\\
\nonumber
&=&\left[\lambda_{in}-\lambda_{ex}\frac{({\bf M}(t)\cdot {\bf H})}{\Omega}\right]\,[{\bf M}(t)\times[{\bf M}(t)\times {\bf H}]],
\end{eqnarray}
entering the basic equation of micromagnetic dynamics (\ref{e1.1}) is the following Modified Landau-Lifshitz (MLL) 
equation 
\begin{eqnarray}
\label{e2.11}
{\dot {\bf M}}=&-&\gamma\mu_0[{\bf M}\times {\bf H}]\\
\nonumber
&-&\left[\alpha\frac{\gamma\mu_0}{M_s}-\beta\left(\frac{\gamma\mu_0}{M_s}\right)^2\frac{({\bf M}\cdot {\bf H})}{\Omega}\right]\,[{\bf M}\times[{\bf M}\times {\bf H}]],
\end{eqnarray}
which obeys all constraints of the canonical LL equation. One sees that unlike the linear-in-magnetic-field intrinsic damping torque, the extrinsic damping torque is described by quadratic-in-magnetic-field relaxation function. At this point it seems noteworthy that the need in allowing for the quadratic-in-$H$
damping terms has been discussed long ago\cite{KP-75}. The above scheme can be regarded, therefore, 
as a development of this line of theoretical investigations.  
In terms of the unit vector of magnetization  ${\bf m}(t)={\bf M}(t)/{M_s}$ and Larmor frequency $\mbox{\boldmath ${\omega}$}=\gamma\mu_0 {\bf H}$ the last 
equation can be converted to (see\cite{S-84} for comparison)   
 \begin{eqnarray}
\label{e2.12}
{\dot {\bf m}}=[\mbox{\boldmath ${\omega}$}\times{\bf m}]-\left[\alpha-\beta\,\frac{({\bf m}\cdot\mbox{\boldmath ${\omega}$})}{\Omega}\right][{\bf m}\times[{\bf m}\times\mbox{\boldmath ${\omega}$}]].
\end{eqnarray} 
It can be seen that the obtained MLL equations (\ref{e2.11}) and (\ref{e2.12}) are reduced to the standard LL equation when the effect of extrinsic stresses is ignored (i.e. $\beta=0$).

\section{Variation method of computing relaxation time and FMR linewidth}
The relaxation time is amongst the primary targets of current FMR experiments. In this section, we present variational method
of analytic computation of the FMR linewidth which is quite different from the well-known solution of the susceptibility solution of LL equation\cite{B-12}. 
At the base of the variation method under consideration lies the equation of the magnetization energy loss from which the exponential relaxation time  
$\tau$ as a function of FMR frequency $f=\omega/(2\pi)$ is derived. 
The FMR linewidth, $\Delta \omega=\gamma\mu_0 \Delta H$, is computed from  the well-known Gabor uncertainty relation (e.g. \cite{B-83}, Sec.11.2)
\begin{eqnarray}
\label{e3.1}
&& \Delta \omega\, \tau =1
\end{eqnarray}
between the full-width-at-half-maximum in the resonance-shaped spectral line $\Delta \omega$ and lifetime $\tau$ of resonance excitation. 
\subsection{FMR linewidth caused by intrinsic damping torque}
For the former we consider relaxation process brought about by intrinsic damping torque.
Our approach is based on the observation that the equation of magnetization energy loss in the process of a uniform precession of magnetization 
in a dc magnetic field 
\begin{eqnarray}
\label{e3.2}
&& \frac{dW_m}{dt}=-\mu_0 H M_s \frac{d(\cos\theta(t))}{dt}=\mu_0 H_i R^{in}_i(t),\\
\label{e3.3}
&& R_i^{in}=\frac{2\lambda_{in}}{\mu\,M_s^2}\gamma_{ik}\,M_l\,\sigma^{in}_{kl},
\end{eqnarray}
is reduced to the equation for the cosine function $u(t)=\cos\theta(t)$ of angle $\theta(t)$ between ${\bf M}(t)$ and ${\bf H}$, namely  
\begin{eqnarray}
\label{e3.4}
\frac{du(t)}{dt}=-\alpha\omega[u^2(t)-1],\quad \omega=\gamma\mu_0 H.
\end{eqnarray} 
The  right hand side of (\ref{e3.4}) suggests that there are two equilibrium configurations, namely, with $u(0)=1$ corresponding 
to ${\bf M}\uparrow\uparrow{\bf H}$ and  $u(\pi)=-1$ corresponding to ${\bf M}\uparrow\downarrow{\bf H}$.   
The stability of these configurations can be assessed by the standard procedure of introducing small-amplitude deviations $\delta u(t)$ from the equilibrium values $u(0)=u_0=\pm 1$. On substituting  $u(t)=u_0+\delta u(t)$,     
into (\ref{e3.4}) with $u_0=1$ and retaining first order terms in $\delta u(t)$ we obtain equations 
describing exponential relaxation of magnetization to the state of stable magnetic equilibrium:    
\begin{eqnarray}
\label{e3.5}
&&\frac{d\delta u(t)}{dt}=-(2\alpha\omega)\delta u\,\to\,\delta u(t)=\delta u(0){\rm e}^{-t/\tau},\\
\label{e3.6}
&& \tau^{-1}=2\alpha\omega,\quad \omega=\gamma\mu_0 H.
\end{eqnarray} 
The second stationary state, with $u_0=-1$, is unstable, since in this case the resultant  
linearized equation, $\delta {\dot u}=(2\alpha\omega)\delta u$, having the solution, $\delta u(t)=\delta u(0){\rm e}^{t/\tau}$, 
describes a non-physical behavior of $\delta u$ as the time is increased.   
Inserting (\ref{e3.6}) in (\ref{e3.1}), we arrive at the basic prediction of the standard micromagnetic model 
 \begin{eqnarray}
 \label{e3.7}
 \Delta H(f)=A\,f,\quad A=\frac{4\pi\alpha}{\gamma\mu_0}.
 \end{eqnarray}
This last equation provides a basis for discussion of empirical linewidth-frequency dependence $\Delta H_{exp}=\Delta H_0+\Delta H_{exp}(f)$ with 
$\Delta H_{exp}(f)=A_{exp}\,f$. Central to such a discussion is the identification of theoretical and experimental linewidths, $\Delta H(f)=\Delta H_{exp}(f)$, from which the magnitude of $\alpha$ is deduced and applied to (\ref{e3.6}) for obtaining numerical estimates of the relaxation time $\tau$.

\subsection{FMR linewidth caused by both intrinsic and extrinsic damping torques}  
In this case the starting point is the equation of magnetization energy loss with the combined relaxation function 
\begin{eqnarray}
\label{e3.8}
&& \frac{dW_m}{dt}=-\mu_0 H M_s \frac{d(\cos\theta(t))}{dt}=\mu_0 H_i R_i(t),\\
\label{e3.9}
&& R_i=\frac{2\lambda_{in}}{\mu\,M_s^2}\gamma_{ik}\,M_l\,\sigma^{in}_{kl}-\frac{\lambda_{\rm ex}}{\mu\Omega}\gamma_{ik}M_l\sigma^{ex}_{lk},
\end{eqnarray} 
which after some algebra is converted into equation for $u$ having the form 
\begin{eqnarray}
\label{e3.10}
\frac{du(t)}{dt}=-\left[\alpha\omega-\beta\frac{\omega^2}{\Omega} u(t)\right] (u^2(t)-1).
\end{eqnarray} 
The right hand side of this equation suggests that there are three stationary state characterized by   
\begin{eqnarray}
\label{e3.11}
u_0(\theta=0)=\pm 1,\,\,u_0(\theta=\theta_M)=\frac{\alpha}{\beta}\frac{\Omega}{\omega}. 
\end{eqnarray} 
Applying to (\ref{e3.10}) the standard linearization procedure $u(t)=u_0+\delta u(t)$ in (\ref{e3.10}) one finds
that resultant equation is equivalent to the equation of exponential relaxation, $\delta {\dot u}=-\tau^{-1}\delta u$, if and only if  
the parameter
\begin{eqnarray}
\label{e3.12}
&& \tau^{-1}=\left[2\left(\alpha\omega -\beta\frac{\omega^2}{\Omega}u_0\right)u_0+\beta\frac{\omega^2}{\Omega}(1-u_0^2)\right]
\end{eqnarray} 
 is a positive constant. It is easy to see that this is the case for $u_0(\theta=0)=1$ and 
 $u_0(\theta=\theta_M)$ given by rightmost of equations (\ref{e3.11}). This latter $u_0$ corresponds to a quasi-stationary transient  
 configuration of precessing magnetization owing its existence to the coupling of magnetization with extrinsic stresses of magnetic anisotropy. 
 The state with $u_0=-1$, is unstable. For the total relaxation time  
 $\tau^{-1}=\tau^{-1}(\theta=0)+\tau^{-1}(\theta=\theta_M)$ and the FMR linewidth (following from Gabor uncertainty relation $\Delta H=[\gamma\mu_0\tau]^{-1}$)
 we obtain 
\begin{eqnarray}
\label{a3.15}
\tau^{-1}&=&2\alpha\omega-\beta\frac{\omega^2}{\Omega}(1+\cos^2\theta_M),\\
\label{a3.16} 
\Delta H&=&\frac{2\omega}{\gamma\mu_0}\left(\alpha -\frac{\beta\omega}{2\Omega}\left[1+\cos^2\theta_M\right]\right)\\
\nonumber
        &=&\frac{4\pi f}{\gamma\mu_0}\left[\alpha -\frac{\beta}{2}\left(1-\left(\frac{\gamma\mu_0 M_s}{2\pi f}\right)^{1/2}\right)\,
        (1+\cos^2\theta_M)\right].
\end{eqnarray} 
It is worth emphasizing that the expounded micromagnetic mechanism of the magnetization precession damping (due to magnetization-stress coupling)
presumes that the process of spin-relaxation is not accompanied by generation of spin-waves (magnons), because the magnetization ${\bf M}$ is regarded as a 
spatially-uniform vector across the multilayered film. At this point the considered  regime of the magnon-free spin relaxation 
(in which the wave vector of spin wave $k=0$) is quite different from spin relaxation caused by two-magnon scattering\cite{AM-99}.   
The most conspicuous feature of this (substantially macroscopic) mechanism, responsible for the 
non-linear frequency dependence of FMR linewidth, is the transient magnetization configuration owing its existence to the 
extrinsic stresses generic to the multilayered films. Such a configuration is absent in perfect monolayered films 
(without impurities and defects of crystalline lattice) of pure ferromagnetic metals (Ni, Co, Fe) whose ferromagnetic properties 
are dominated by intrinsic stresses of magnetic anisotropy.

\begin{figure}
\centering{\includegraphics[width=8.5cm]{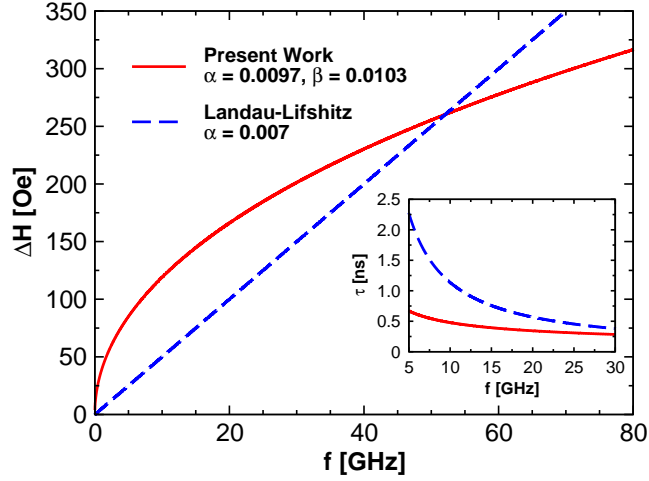}}
\caption{
 The FMR linewidth $\Delta H$ and relaxation time $\tau$ as functions of the FMR frequency $f$, 
 computed on the basis of the standard and modified in the present work Landau-Lifshitz equation.}
\end{figure}

\section{Discussion and summary}
In approaching the interpretation of FMR measurements in terms of presented theory, in the remainder of this work, we focus on a case of in-plane configuration (${\bf M}\uparrow\uparrow{\bf H}$)  which is of particular interest in connection with 
the recent discovery of non-linear frequency dependence of FMR linewidth\cite{B-11,WH-04,L-06}. In this case, the last equation for the FMR linewidth takes 
the form  
\begin{eqnarray}
\label{e3.18}
&& \Delta H=\frac{4\pi f}{\gamma\mu_0}\left[\alpha-\beta(1-({\gamma\mu_0M_s}/{2\pi f})^{1/2})\right].
\end{eqnarray}   

\begin{figure}
\centering{\includegraphics[width=8.0cm]{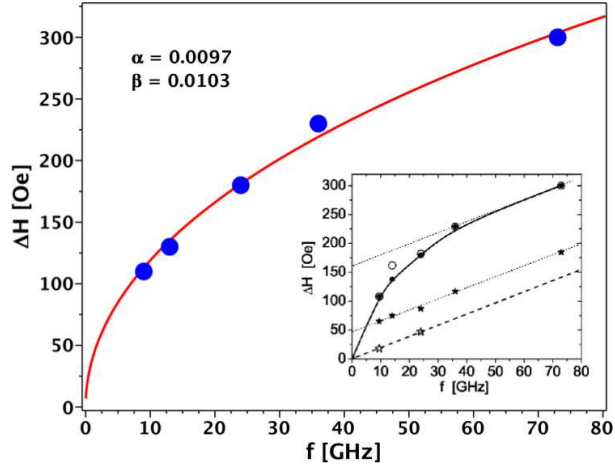}}
\caption{
 Theoretical fit, equation (\ref{e3.18}), of the empirical non-linear frequency dependence of FMR linewidth detected in the  
 FMR measurements\cite{WH-04} on multilayered metallic nanostructures Pd/Fe/GaAs.}
\end{figure}     
  
\begin{figure}
\centering{\includegraphics[width=8.0cm]{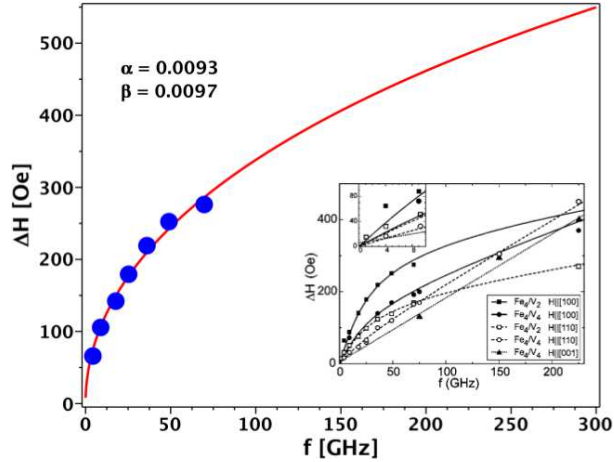}}
\caption{
 Same as Fig.2, but for the thin film of Fe/V  measured in\cite{L-06}.}
\end{figure}   
  
To illuminate the difference between predictions of the standard and modified LL models, in Fig.1 we plot $\tau$ and $\Delta H$ as functions of 
the FMR frequency $f$ computed with the pointed out parameters of $\alpha$ and $\beta$. In computation based 
on the  standard micromagnetic model, equation (\ref{e3.7}), we have used one and the same value of parameter $\alpha$ as in\cite{C-06} reporting 
the FMR measurements on ultrathin films of Permalloy. 
The presented in Fig.1 values of $\alpha$ and $\beta$ have been deduced from fitting, equation (\ref{e3.18}), of the non-linear frequency dependence of FMR linewidth discovered in the FMR measurements\cite{WH-04}. The result of this fit is shown in Fig.2. In Fig.3, we plot our fit of the FMR linewidth measurements
\cite{L-06} on multilayered samples of Fe/V. 
A more detailed discussion of consequences of considered micromagnetic mechanism of spin relaxation will be the subject of forthcoming article.

\end{document}